\documentstyle[aps,prl]{revtex}

\begin{document}
\title{\bf  The Pauli Equation for Probability Distributions}

\author{Stefano Mancini$^1$, 
Olga V. Man'ko$^2$,  
Vladimir I. Man'ko$^2$,
and Paolo Tombesi$^3$}

\address{
${}^1$ INFM, Dipartimento di Fisica,
Universit\`a di Milano,
Via Celoria 16, I-20133 Milano, Italy\\
${}^2$ P.N. Lebedev Physical Institute,
Leninskii Prospekt 53, Moscow 117934, Russia\\
${}^3$INFM, Dipartimento di Matematica e Fisica,
 Universit\'a di Camerino, 
I-62032 Camerino, Italy}

\date{Received: \today}

\maketitle
\widetext

\begin{abstract}
The ``marginal'' distribution for measurable 
coordinate and spin projection is
introduced. Then, the analog of the Pauli 
equation for spin-$\frac{1}{2}$
particle is obtained for such probability 
distributions instead of the usual
wave functions. That allows a classical-like 
approach to quantum mechanics.
Some illuminating examples are presented. 
\end{abstract}


\section{Introduction}

Since the early days of quantum mechanics, we have been 
forced to coexist with
complex probability amplitudes without worrying about 
their lack of any
reasonable physical meaning. One should not ignore, 
however, that the
wave-like properties of quantum objects still raise 
conceptual problems on
whose solutions, a general consensus is far from 
having been 
reached~\cite{epr,erw}.

A possible way out of this difficulty has been 
implicitly suggested by
Feynmann~\cite{fey}, who has shown that, by dropping 
the assumption that the
probability for an event must always be nonnegative, one 
can avoid the use
of probability amplitudes in quantum mechanics. This 
proposal goes back to
the work by Wigner~\cite{wig}, who first introduced 
nonpositive
pseudoprobabilities to represent quantum mechanics in 
phase space, and to
the Moyal approach to quantum mechanics~\cite{moy}.

From a conceptual point of view, the elimination of the 
waves from quantum
theory is in line with the procedure inaugurated by 
Einstein with the
elimination of the aether in the theory of electromagnetism.

The phase-space formulation of quantum 
mechanics~\cite{wig,hus,glasud}
provides a means of analyzing quantum-mechanical 
systems while still
employing a classical framework. Moreover, a 
quantum mechanics without wave
functions has been discussed in~\cite{wan}.

Recentely, the problem of quantum state measurement, 
initially posed 
by Pauli \cite{pauli}, received a lot of 
attention \cite{special}.
The tomographic approach~\cite{ber,vr} to the 
quantum state of a
system has allowed to establish a map between the 
density operator (or any its
representation) and a set of probability 
distributions, often called
``marginals". The latter have all the 
characteristics of classical
probabilities; they are nonnegative, measurable, 
and normalized.

Based on this connection, a classical-like description 
of quantum dynamics by
means of ``symplectic tomography" has been 
formulated \cite{fou}, providing
a bridge between classical and quantum worlds. 
That is, the evolution of a
quantum system with continuous observables (namely, 
quadrature components of
a field mode) was described in terms of a 
classical-like equation for a
marginal distribution. 
Different aspects of this classical-like description 
using tomographic
probabilities were recently analyzed~\cite{rosa,anju}.

On the other hand, discrete observables, like spin or 
angular momentum, are 
so important in quantum mechanics as the continuous ones are. 
Hence, the tomography scheme 
for discrete variables was introduced \cite{ulf}, 
and the marginal
distribution for rotated spin variables has been 
constructed 
\cite{dodman}, deriving an evolution equation for 
this function.

Here, we would extend the approach by considering a 
spin-$\frac{1}{2}$
particle moving in a potential, then constructing 
the marginal
distributions for space coordinates and spin 
projections and finally deriving
the evolution equation for such probabilities, 
which would be the analog 
of the Pauli equation. It would also be a 
generalization of approaches attempted 
by us in previous papers \cite{fou}.

Essentially, our aim is to eliminate the hybrid 
procedure of describing the
dynamical evolution of a system, which consists of 
a first stage, where
the theory provides a deterministic evolution of the 
wave function, followed
by a hand-made construction of the physically 
meaningful probability
distributions. If the probabilistic nature of the 
microscopic phenomena is
fundamental, and not simply due to our ignorance, 
as in classical statistical
mechanics, why should it be impossible to describe 
them in probabilistic
terms from the very beginning? On the other hand, 
the language of
probability, suitably adapted to take into account 
all the relevant
constraints, seems to be the only language capable 
of expressing the
fundamental role of ``chance'' in nature~\cite{kuhn}.

The paper is organized as follow:

In Section~2, we review the general approach to 
construct known tomography 
schemes using density matrix in the specifically 
transformed reference 
frames. In Section~3, we derive the general 
evolution equation for
tomographic probabilities (marginal distributions) 
which describe the quantum
state instead of density matrix. In Section~4, 
the general scheme of
tomography construction is used to re-derive the 
particular example of
symplectic tomography, which is applied for 
measuring states depending on
continuous quadrature. In Section~5, the general 
scheme is used to re-derive 
the construction of spin-state tomography. 
In Section~6, the general scheme 
of Section~2 is then applied to obtain tomographic 
probabilities in the combined
situation described by spatial (multidimensional too)
and spin variables. In Section~7, some examples are 
studied in the context of the probability
representation of quantum mechanics. Section~8 concludes.

We are using the natural unit ($\hbar=c=1$).

\section{General approach to quantum tomography}

In this section, we give a short review of the general 
principles used to 
construct a tomography scheme for measuring quantum states.
Recently, we established ~\cite{jmo} a quite general 
principle
of constructing measurable probabilities, which 
determine completely the quantum state
in the tomographic approach; more refined treatments 
then followed
\cite{brief,dar}. 
Here, we apply our general approach to derive 
the evolution equation 
for the tomographic probabilities that is 
alternative in some sense 
to the Schr\"odinger equation 
for the wave function (or the quantum Liouville 
equation for the density
matrix). 

Let us consider a quantum state described by the density
operator ${\hat\rho}$, which is a nonnegative 
hermitian operator, i.e., 
\begin{equation}  \label{rho}
{\hat\rho}^{\dag}={\hat\rho}, \qquad \mbox {Tr}
\,{\hat\rho}=1\,,
\end{equation}
and 
\begin{equation}  \label{rhovv}
\langle v\mid{\hat\rho}\mid v \rangle=\rho_{v,v}\geq 0.
\end{equation}
We label the vector basis $\mid v \rangle $ in the 
space of pure quantum states
by the multidimensional index $v=\left(v_1,v_2,\ldots,v_N
\right),$
where the number $N$ shows the number of degrees of 
freedom of the system
under consideration. Among indexes $v_k$, $~k=1,\ldots,N$, 
there are
continuous ones like position (or momentum) and 
discrete ones like  spin
projections. In this sense, the wave function 
$\psi(v)=\langle v \mid\psi\rangle$ of a pure state 
$\mid\psi\rangle$ depends both on continuous
and discrete observables. Formula~(\ref{rhovv}) can be 
rewritten by using the
hermitian projection operator 
\begin{equation}  \label{Piv}
{\hat \Pi}_{v}=\mid v \rangle\langle v \mid,
\end{equation}
in the following form 
\begin{equation}  \label{rhovvPi}
\rho_{v,v}=\mbox{Tr}
\left\{{\hat \Pi}_{v}{\hat\rho}\right\}\,.
\end{equation}

The physical meaning of the projector ${\hat \Pi}_{v}$ 
is that it extracts the
state $\mid v \rangle$ with given $v$ (for example, 
with given
position and spin projection), which is an eigenstate of 
the-commuting-hermitian operators ${\hat V}
=\left({\hat V}_1,{\hat V}_2,\ldots,
{\hat V}_N\right)$
\begin{equation}  \label{Vk}
{\hat V}_k\mid v \rangle=v_k\mid v \rangle .
\end{equation}
In the space of states, there is a family of unitary 
transformation operators 
${\hat U}(\sigma)$ depending on the parameters 
$\sigma=(\sigma_1,\ldots,\sigma_k\ldots)$, 
that can be sometimes identified with a 
group-representation operators. In
these cases, the parameters $\sigma$ describe 
the group element. 
It was shown \cite{jmo,my} that known tomography schemes 
can be
considered from the viewpoint of the group theory 
by using appropriate groups. 
More recentely this concept has been developed 
obtaining an elegant group theoretical 
approach to quantum state measurement \cite{dar}.
Here, we formulate the tomographic approach in the 
following way. Let us
introduce a ``transformed density operator''
\begin{equation}  \label{rhosig}
{\hat\rho}_{\sigma}={\hat U}^{-1}(\sigma){\hat\rho} 
{\hat U}(\sigma).
\end{equation}
Its diagonal elements are still nonnegative probabilities 
\begin{equation}  \label{wz}
\langle z \mid{\hat\rho}_{\sigma} \mid z \rangle
=\langle\langle z \mid{\hat\rho}\mid z \rangle\rangle
\equiv w(z,\sigma)\,.
\end{equation}
Here, $\mid z\rangle$ is one of the possible vectors 
$\mid v\rangle$,
while the symbol $\mid z\rangle\rangle$ denotes the 
transformed vectors
\begin{equation}\label{zvec}
\mid z\rangle\rangle={\hat U}(\sigma) \mid z\rangle\,,
\end{equation}
which in turn 
are eigenstates of the-transformed-operators
\begin{equation}\label{Zop}
{\hat Z}={\hat U}(\sigma){\hat V}{\hat U}^{-1}(\sigma)\,.
\end{equation}
As a consequence of the unit trace of the density 
operator we also have 
the normalization condition 
\begin{equation}  \label{normz}
\int \,dz\,w\left(z,\sigma\right)=1.
\end{equation}
Of course, in case of discrete indices, the integral 
in Eq.~(\ref{normz}) 
is replaced by a sum over discrete variables.

Formula~(\ref{wz}) can be interpreted as the probability 
density for the measurement of the observable ${\hat V}$ 
in an ensemble of
transformed reference frames labeled by the index $\sigma$, 
if the state
${\hat\rho}$ is given. Along with this interpretation, 
one can also
consider the transformed projector
\begin{equation}\label{Pizsig}
{\hat \Pi}_z(\sigma)
={\hat U}(\sigma){\hat\Pi}_z {\hat U}^{-1}(\sigma)
=\mid z\rangle\rangle\langle\langle z \mid\,,
\end{equation}
the explicit expression for the
probability $w \left(z,\sigma\right)$ takes the form 
\begin{equation}\label{wzPi}
w\left(z,\sigma\right)=
\mbox{Tr}\left\{ {\hat\rho}{\hat\Pi}_z(\sigma) \right\}
=\mbox{Tr}\left\{{\hat\rho} 
\mid z\rangle\rangle\langle\langle z\mid
\right\}\,.  
\end{equation}
These probability densities are also called 
``marginal" distributions 
as generalization of the concept introduced by 
Wigner \cite{wig}.
The tomography schemes are 
based on the possibility to find the inverse of 
Eq.~(\ref{wzPi}). If it is
possible to solve Eq.~(\ref{wzPi}), considering 
the probability 
$w\left(y,\sigma \right)$ as known function and 
the density matrix as unknown
operator, the quantum state can be described by 
the positive probability
instead of the density matrix. This property is 
the essense of state reconstruction 
techniques.
In such cases, the inverse of Eq.~(\ref{wzPi}) 
takes the form 
\begin{equation}\label{rhoK}
{\hat\rho} =\int w\left (z,\sigma\right){\hat K}
\left (z,\sigma \right)~dz~d\sigma.  
\end{equation}
Thus, there exist a family of operators 
${\hat K}\left(z,\sigma\right)$ 
depending on both the variables $z$ and parameters 
$\sigma$ such that the density operator is reconstructed, 
if the probability 
$w \left(z, \sigma\right)$ is known. 
It is worth remarking that 
transformations ${\hat U}(\sigma)$ can form other 
algebraic constructions, 
which have no structure of groups \cite{my}. 
The only condition 
for the existence of a tomography scheme is 
the possibility to invert 
Eq.~(\ref{wzPi}). In the cases of optical 
tomography~\cite{vr}, 
symplectic tromography~\cite{fou}, and spin 
tomography~\cite{dodman,olga},
the sets of transformations ${\hat U}(\sigma)$ 
have the structure of
corresponding Lie groups (i.e., rotation, 
symplectic and spin).

\section{The time evolution equation}

We are now interested in finding the evolution equation for the
probability $w\left (z,\sigma,t\right)$, 
in which $t$ is the time
parameter. Using Eq.~(\ref{wzPi}) one has 
\begin{equation}\label{dtw}
\partial_t\, w\left(z,\sigma,t\right)=
\mbox {Tr}\left\{ \left[\partial_t\,{\hat\rho}(t)\right]
{\hat\Pi}_z(\sigma)
\right\}\,.  
\end{equation}

On the other hand, the density operator satisfies the 
Liouville-Von Neumann equation 
\begin{equation}\label{LVNeq}
{\partial_t\,{\hat\rho}}(t)=i\left[{\hat\rho} (t),
{\hat H}\right]\,, 
\end{equation}
with ${\hat H}$ the system Hamiltonian.
By inserting Eq.(\ref{LVNeq}) in (\ref{dtw}), 
and with the aid of Eq.(\ref{rhoK}), we find the evolution 
equation for the probability 
$w$ in a closed form 
\begin{equation}\label{inteq}
\partial_t\, w\left(z,\sigma,t\right)
=\int dz'~d\sigma'~w\left(z',\sigma',t\right) 
\mbox{Tr}\left\{i\left[{\hat K}(z',\sigma'),
{\hat H}\right]{\hat\Pi}_z(\sigma)\right\}\,. 
\end{equation}
Equation~(\ref{inteq}) represents the 
classical-like version of the 
Liouville-Von Neumann equation, thus, it would be 
the analog of the Pauli 
equation for a system with space and spin degrees 
of freedom.

\section{Quantum tomography with continuous variables}

Let us consider, in a one-dimensional system,  
an operator ${\hat X}$ as the linear 
combination of position ${\hat q}$ and
momentum ${\hat p}$~\cite{qso95,qso96}
\begin{equation}\label{quadef}
{\hat X}=\mu{\hat q}+\nu{\hat p}\,,
\end{equation}
which depends upon real parameters $\mu$, $\nu$ and, due to 
its hermiticity, is a measurable
observable. Since the linear canonical 
transformation~(\ref{quadef})
belongs to the symplectic group $Sp(2,R)$, 
the tomography scheme under discussion
was called ``symplectic tomography''~\cite{qso96}.

The probability (marginal) related to the 
observable (\ref{quadef})
is given by
\begin{equation}\label{wx}
w(x,\mu,\nu)=\langle\langle x\mid\hat\rho
\mid x\rangle\rangle\,,
\end{equation}
where ${\hat\rho}$ is the system's density operator,
while the eigenstates $\mid x\rangle\rangle$ of 
the operator (\ref{quadef}) can be 
written as
\begin{equation}\label{xvec}
\mid x\rangle\rangle=\int dq\,
\langle q\mid x\rangle\rangle\,\mid q\rangle\,,
\end{equation}
with $\mid q\rangle$ the position eigenkets.
The wave function $\langle q\mid x \rangle\rangle$ 
can be easily calculated
by using the following equality
\begin{equation}\label{equality}
\langle q\mid {\hat X} \mid x \rangle\rangle=
\langle q\mid \mu {\hat q}+\nu {\hat p} 
\mid x \rangle\rangle\,,
\end{equation}
and then transforming it in a partial 
differential equation
\begin{equation}\label{wfeq}
x\,\langle q\mid x\rangle\rangle 
=\mu q\,\langle q\mid x \rangle\rangle
-i\nu\frac{\partial}{\partial q}\langle 
q\mid x \rangle\rangle\,.
\end{equation}
The solution is 
\begin{equation}\label{wf}
\langle q\mid x \rangle\rangle
=\left(\frac{1}{|\nu|}\right)^{1/2}
\exp\left[i\frac{x}{\nu}q-\frac{i}{2}
\frac{\mu}{\nu}q^2\right]\,.
\end{equation}
It is worth noting that as soon as $\mu\to 1$ 
and $\nu\to 0$, then 
$\mid x\rangle\rangle\to\mid x\rangle$ and the 
wavefunction (\ref{wf})
tends to $\delta(q-x)$.

Furthermore, Eq.(\ref{wx}) can be formally rewritten as
\begin{equation}\label{wxPi}
w(x,\mu,\nu)={\rm Tr}\left\{{\hat\rho}
{\hat\Pi}_x(\mu,\nu)\right\}\,,
\end{equation} 
where the transformed projector is given by
\begin{equation}\label{Pix}
{\hat \Pi}_x(\mu,\nu)={\hat U}(\mu,\nu)
{\hat \Pi}_x{\hat U}^{-1}(\mu,\nu)\,,\quad
{\hat \Pi}_x=\mid x\rangle \langle x\mid\,.
\end{equation}
Here, the transformation ${\hat U}(\sigma)$ 
is chosen to be the
symplectic group representation~\cite{jmo}
\begin{equation}\label{Uspace}
{\hat U}(\mu,\nu)
=\exp \left[i\phi \left(\frac {\hat p^2}{2}
+\frac {\hat q^2}{2}
\right)\right]\exp\left[\frac {i\lambda}{2}
\left(\hat q \hat p+
\hat p \hat q\right)\right]\,.
\end{equation}
The rotation and scaling 
parameters $\phi$ and $\lambda$ are related to 
$\mu$ and $\nu$ by the following
formulas
\begin{eqnarray}
\mu=\lambda \cos \phi\,,&\quad &\nu
=\lambda^{-1}\sin \phi\,,
\nonumber\\[-2mm]
&&\label{param}\\[-2mm]
\phi =\frac{1}{2}\,\mbox{arcsin}
\left(2\mu \nu\right),&\quad &
\lambda=\pm\frac{1}{4}\,
\sqrt{\frac{1\pm\sqrt{1-4\mu^2\nu^2}}{2}}\,.
\nonumber
\end{eqnarray}
This means that the marginal distribution 
$w\left(x,\,\mu,\,\nu\right)$
for this particular case of symplectic 
tomography is given by the relation
\begin{eqnarray}\label{wxTr}
w\left(x,\mu,\nu\right)&=&\mbox{Tr}\,
\left\{\mid x\rangle \langle x \mid
\exp\left[i\phi\left(\frac{\hat p^2}{2}
+\frac {\hat q^2}{2}\right)\right]
\exp \left[\frac{i\lambda}{2}\left(
\hat q\hat p+\hat p\hat q\right)
\right]\right.\nonumber\\
&&\qquad\times\left. {\hat\rho} \exp\left[
-i\phi\left(\frac{\hat p^2}{2}
+\frac {\hat q^2}{2}\right)\right]
\exp\left[-\frac{i\lambda}{2}\left(\hat q\hat p
+\hat p\hat q\right)\right]
\right\}.
\end{eqnarray}
Such measurable probability can be explicitly 
expressed as \cite{qso95}
\begin{equation}\label{wxint}
w(x,\mu,\nu)=\int\,dy\,dk\,\exp\left[
-ikx+\frac{i\mu\nu k^2}{2}
+iky\mu\right]
\rho(y+\nu k,y)\,,
\end{equation}
where $\rho(y+\nu k,y)=\langle y
+\nu k|{\hat\rho}|y\rangle$ 
is the representation of the density
matrix over the position eigenkets.
The marginal satisfy the following homogeneous property
\begin{eqnarray}
w(x,\mu,\kappa\nu)&=&\frac{1}{\kappa}\,w(x/\kappa,
\mu/\kappa,\nu)\,,
\nonumber\\[-2mm]
\label{hom}\\[-2mm]
w(x,\kappa\mu,\nu)&=&\frac{1}{\kappa}\,w(x/\kappa,
\mu,\nu/\kappa)\,.
\nonumber
\end{eqnarray}

The above relation~(\ref{wxint}) can be 
inverted~\cite{qso96} as
\begin{equation}\label{rhoKsym}
{\hat\rho}=\int\,dx\,d\mu\,d\nu\;w(x,\mu,\nu)\,
{\hat K}(x,\mu,\nu)\,,
\end{equation}
where the kernel operator takes the form
\begin{equation}\label{Ksym}
{\hat K}(x,\mu,\nu)=\frac{1}{2\pi}\epsilon^2\exp\left[
-i\epsilon X+\frac{i\epsilon^2\mu\nu}{2}\right]\,
e^{i\epsilon\mu{\hat q}}\,e^{i\epsilon\nu{\hat p}}\,.
\end{equation}
Here, $\epsilon$ can be set equal 1; this freedom reflects the 
overcompleteness of information obtainable
by means of all possible 
marginals~(\ref{wxTr})~\cite{qso95,qso96}.

The multi-mode generalization~\cite{qso96} is straightforward, 
and the analog of formula~(\ref{wxTr}) holds with the 
following replacement
\begin{eqnarray}
\mid x\rangle &\longrightarrow &\mid \vec{x}\,\rangle ,\quad 
\vec{x}=\left(x_1,x_2,\ldots\right),\nonumber\\
\phi\left(\frac {\hat p^2}{2}+\frac {\hat q^2}{2}\right)&
\longrightarrow &
\phi_1\left(\frac {\hat p_1^2}{2}+\frac {\hat q_1^2}{2}\right)+
\phi_2\left(\frac {\hat p_2^2}{2}+\frac {\hat q_2^2}{2}\right)+
\ldots,\\
\lambda \left(\hat q\hat p+\hat p\hat q\right)&\longrightarrow &
\lambda_1 \left(\hat q_1\hat p_1+\hat p_1\hat q_1\right)+
\lambda_2 \left(\hat q_2\hat p_2+\hat p_2\hat q_2\right)+
\ldots.\nonumber
\end{eqnarray}
Relations of the parameters $\lambda_k, ~\phi_k$ 
to the parameters $\mu_k, ~\nu _k$ are the same of 
Eq.~(\ref{param}).

\section{Quantum tomography with discrete variables}

Here, we consider a spin-$j$ system.
Following~\cite{dodman,olga} we will derive 
the expression for the
density matrix of a spin state in terms of measurable 
probability
distributions. 

For arbitrary values of spin, let the spin state 
have the density matrix 
\begin{equation}\label{rhommp}
\rho _{mm'}^{(j)}=\langle jm\mid  
{\hat\rho}^{(j)}\mid jm'\rangle \,,
\qquad m=-j,-j+1,\ldots,j-1,j\,, 
\end{equation}
where
\begin{equation}\label{spinop}
\hat j_3 \mid j m \rangle = m \mid j m \rangle\,,\quad
\hat j^2 \mid j m\rangle = j(j+1) \mid j m\rangle\,,
\end{equation}
and
\begin{equation}\label{rhoj}
{\hat\rho}^{(j)}=\sum_{m=-j}^j \sum_{m'=-j}^j 
\rho_{mm'}^{(j)}\mid j m\rangle\langle j m'\mid \,.
\end{equation}
The operator ${\hat\rho}^{(j)}$ is the density operator 
of the state under 
consideration.

The general group construction of tomographic
schemes \cite{jmo} was also used for spin tomography
\cite{dodman,olga}. 
The idea is to consider the diagonal elements of
the density matrix ${\hat\rho}$ in another reference frame,
i.e. rotated one.
To this end we introduce a rotated measureble spin projection
\begin{equation}\label{J3}
{\hat J}_3(\alpha,\beta,\gamma)={\hat D}(\alpha,\beta,\gamma)
{\hat j}_3{\hat D}^{-1}(\alpha,\beta,\gamma)\,,
\end{equation}
where the unitary rotation operator ${\hat D}$ 
depends on the Euler angles
$\alpha ,\,\beta ,\,\gamma \,$. 
The role of the observable ${\hat Z}$ is now 
played by the spin projection ${\hat J}_3$,
while the rotation-transformation parameters 
are the Euler angles $\sigma_1=\alpha,$ 
$\sigma_2=\beta$, $\sigma_3=\gamma$.
The transformation ${\hat U}(\sigma)$ is 
given by the matrix representation 
of the rotation group, i.e., the Wigner 
$D$-function \cite{bie}.

The marginals are
\begin{equation}\label{ws}
w\left(s,\alpha,\beta,\gamma\right)=
\langle\langle j s\mid {\hat\rho}\mid j s \rangle\rangle\,,
\end{equation}
where the rotated spin states becomes
\begin{equation}\label{svec}
\mid js \rangle\rangle=\sum_{m=-j}^{j}
D^{(j)\,*}_{s\,m}(\alpha,\beta,\gamma)\mid jm\rangle\,.
\end{equation}
Here the matrix elements 
$D_{m'\,m}^{(j)}
\left (\alpha, \beta, \gamma \right)$~(Wigner 
$D$-functions) are the matrix elements of the 
rotation-group representation \cite{bie}
\begin{equation}\label{D}
D^{(j)}_{m'm}(\alpha,\beta,\gamma)
=e^{i m'\gamma}\,d_{m'm}^{(j)}(\beta)\,
e^{i m \alpha}\,,
\end{equation}
where 
\begin{equation}\label{d}
d_{m'm}^{(j)}(\beta)= 
\left[\frac{(j+m')!(j-m')!}{(j+m)!(j-m)!}\right]^{1/2}
\left(\cos\,\frac{\beta}{2}\right)^{m'+m} \left(\sin\,
\frac{\beta}{2}\right)^{m'-m} P_{j-m'}^{(m'-m,m'+m)}
(\cos\,\beta)\,,
\end{equation}
with $P_n^{(a,b)}(x)$ the Jacobi polynomials \cite{bie}.

Moreover, the transformed spin projector will be
\begin{equation}\label{Pis}
{\hat \Pi}_s(\alpha,\beta,\gamma)
={\hat D}(\alpha,\beta,\gamma)\mid j s\rangle\langle j s\mid 
{\hat D}^{-1}(\alpha,\beta,\gamma)
=\mid js \rangle\rangle 
\langle\langle js\mid\,,
\end{equation}
then, we have
\begin{equation}\label{wsex}
w\left(s,\alpha,\beta,\gamma\right) = 
\sum^j_{m_1=-j}\,\sum^j_{m_2=-j}
\,D_{s\,m_1}^{(j)}(\alpha, \beta,\gamma)\,
\rho^{(j)}_{m_1\,m_2}\,D_{s\,m_2}^{(j)\,*}
(\alpha, \beta, \gamma)\,.
\end{equation}
Since
\begin{equation}\label{Dstar}
D_{m'm}^{(j)\ast}(\alpha,\beta,\gamma)=(-1)^{m'-m}D_{-m'
-m}^{(j)}(\alpha,\beta,\gamma)\,,
\end{equation}
the marginal distribution really depends only on two angles, 
$\alpha$ and 
$\beta$.
Hence
\begin{equation}\label{wsto}
w\left(s,\alpha,\beta,\gamma\right)
\to w\left(s,\alpha,\beta\right)
\,,
\end{equation}
which satisfies the normalization condition
\begin{equation}\label{wsnorm}
\sum_{s=-j}^jw\left(s,\alpha,\beta\right)=1\,.
\end{equation}

As an example, for a spin-$\frac{1}{2}$ state with spin 
projection $+1/2$, we have 
\begin{equation}\label{example}
{\hat\rho}=\pmatrix{1 & 0 \cr 0 & 0}\,,
\end{equation}
and the marginal distributions will be
\begin{equation}\label{wspm}
w\left (s=\mbox{$\frac {1}{2}$},\alpha,\beta\right )
=\cos ^2\frac {\beta}{2}\,,
\qquad 
w\left (s=-\mbox{$\frac {1}{2}$},\alpha,\beta\right )
=\sin ^2\frac {\beta}{2}\,.
\end{equation}
In Refs.~\cite{dodman,olga}, in view of the properties of 
the Wigner $D$-function and the 
Clebsch--Gordan coefficients, Eq.~(\ref{wsex}) was 
inverted and the
density matrix was expressed in terms of the marginal 
distribution
\begin{eqnarray}\label{rhows}
\rho_{m_1\,m_2}^{(j)}&=&  
(-1)^{m_2}\sum_{j_3=0}^{2j}\,\sum_{m_3=-j_3}^{j_3}\,(2j_3+1)^2
\sum_{s=-j}^{j} \int (-1)^{s} w\left(s,\alpha,\beta\right)\, 
\nonumber\\
&\times& 
D_{0\,m_3}^{(j_3)}(\alpha,\beta,\gamma)
W^{j\,j\,j_3}_{s\,-s\,0}\,W^{j\,j\,j_3}_{m_1\,-m_2\,m_3}
\,\frac{d\Omega}{8\pi^2}
\end{eqnarray}
where $m_1,m_2=-j,-j+1,\ldots ,j$ and 
$W^{j_1\,j_2\,j_3}_{m_1\,m_2\,m_3}$ 
are the Wigner-$3\,j$ symbols
\cite{bie}.  The integration is performed over 
the rotation parameters, i.e.
\begin{equation}\label{solid}
\int\,d\Omega\,=\int_0^{2\pi}\,d\alpha\,
\int_0^{\pi}\,\sin\beta\,d\beta\,\int_0^{2\pi}\,d\gamma\,.
\end{equation}
Equation~(\ref{rhows}) can be presented in an invariant 
operator 
form~\cite{olga}. We systematically introduce the 
following notation, 
first for the function on the unit sphere
\begin{equation}\label{Phi}
\Phi_{j\,m_1\,m_2}^{(j_3)}(\alpha, \beta) 
=(-1)^{m_2}\sum_{m_3=-j_3}^{j_3}
D_{0\,m_3}^{(j_3)} (\alpha,\beta,\gamma)
W^{j\,j\,j_3}_{m_1\,-m_2\,m_3},
\end{equation}
and then for the operator on the unit sphere
\begin{equation}\label{Aj}
\hat A_j^{(j_3)}(\alpha,\beta)=(2j_3+1)^2
\sum_{m_1=-j}^{j}\,\sum_{m_2=-j}^j
\mid j m_1\rangle \, \Phi_{j\,m_1\,m_2}^{(j_3)} 
(\alpha,\beta) \, \langle j m_2\mid\,.
\end{equation}
In order to write a final expression for the 
density operator, we introduce 
an operator on the unit sphere which contains 
a dependence on the measurable 
projection of the spin
\begin{equation}\label{Kj}
{\hat K}^{(j)}(s,\alpha,\beta)=(-1)^{s}\sum_{j_3=0}^{2j}
W^{j\,j\,j_3}_{s\,-s\,0} \, {\hat A}_j^{(j_3)}(\alpha,\beta)\,.
\end{equation}
Finally, we obtain a compact expression for the density operator,
\begin{equation}\label{rhoKj}
\hat\rho^{(j)}=\sum_{s=-j}^j
\int\frac{d \Omega}{8\pi^2}\,w(s,\alpha,\beta)\, 
{\hat K}^{(j)}(s,\alpha,\beta).
\end{equation}
Formula~(\ref{rhoKj}) admits of the following 
interpretation. 
To determine the spin state for a spin $j$, 
one has to experimentally measure the
projection $s$ of the spin for each direction 
specified by the angles 
$\alpha$ and $\beta$, obtaining a
distribution function $w\left(s,\alpha,\beta\right)$.
The sum on the r.h.s. of Eq.(\ref{rhoKj}) for a given 
point on the unit 
sphere represents the average operator
$\langle \hat K^{(j)}\left(s,\alpha,\beta\right)\rangle$. 
Then, 
the integral over the whole solid angle gives the 
desired density operator.
Finally, we recognize that, for the spin case, the 
operator~(\ref{Kj}) plays the role 
of the operator ${\hat K} \left(z,\sigma\right)$ 
of Eq.~(\ref{rhoK}), 
employed in the general scheme of Section~2.

\section{The general case}

We are now able to consider the case of
a particle with $N-1$ spatial degrees of freedom, plus one 
spin-$\frac{1}{2}$ degree. 
In this case, 
the state vector $\mid v \rangle$ has the form
\begin{equation}\label{vtot}
\mid {\vec q},m\rangle=
\mid q_1,\ldots q_{N-1}\rangle\otimes\mid
\mbox{$\frac{1}{2}$},\,
m\rangle\,,
\end{equation}
where ${\vec q}$ is the eigenvalue of the position
operator $\hat{\vec q}$ and
the spin projection $m=\left(-1/2,\,1/2\right)$ 
is the eigenvalue 
of the Pauli matrix ${\hat\sigma}_z$.

The transformation operator ${\hat U}(\sigma)$ used to 
construct the tomography 
scheme, for this case, depends on $2(N-1)$ parameters
determining the
symplectic transform, and on
three 
Euler angles determining the spin rotation.

The transformation operator ${\hat U}(\sigma)$ of 
Eq.~(\ref{rhosig}) becomes
the product of operators
\begin{equation}\label{Utot}
{\hat U}(\sigma)=\otimes_{k=1}^{N-1} 
{\hat U}\left(\mu_k,\,\nu_k\right)
\otimes
{\hat U}\left(\alpha,\beta,\gamma\right).
\end{equation}

For the case of spin-$\frac{1}{2}$,
the representation of the rotation group is given by
\begin{equation}\label{Dmatrix}
D(\alpha,\beta,\gamma)=
\left(\begin{array}{cc}
e^{i\alpha/2}\cos\left(\beta/2\right)e^{-i\gamma/2}&
-e^{-i\alpha/2}\sin\left(\beta/2\right)e^{i\gamma/2}\\[2mm]
e^{i\alpha/2}\sin\left(\beta/2\right)e^{-i\gamma/2}&
e^{i\alpha/2}\cos\left(\beta/2\right)e^{i\gamma/2}
\end{array}\right),
\end{equation}
which determines the operator
\begin{equation}\label{Uspin}
{\hat U}\left(\alpha,\beta,\gamma\right)=
\sum_{m_1=-1/2}^{1/2}~
\sum_{m_2=-1/2}^{1/2}D^{(1/2)}_{m_1\,m_2}
\left(\alpha,\beta,\gamma\right)
\mid \mbox{$\frac{1}{2}$},\,m_1\rangle\langle 
\mbox{$\frac{1}{2}$},\,m_2\mid\,.
\end{equation}

The marginal distribution 
$w\left(z,\sigma\right)$~(\ref{wzPi}) 
will depend on $N-1$ continuous
(noncompact) variables $z_1=x_1,$ $\ldots,$ 
$z_{N-1}=x_{N-1},$ and one discrete spin projection 
$z_N=s,$ as well as on parameters
$\mu_k,\,\nu_k$ and on Euler angles 
$\alpha,\,\beta.$ The dependence of the marginal 
distribution on 
the Euler angle $\gamma$ disappears, as it was shown in the 
previous section, due to the structure of Wigner $D$-functions.

In order to get an analog of the Pauli evolution 
equation for the
marginal distribution, we consider the general 
equation~({\ref{inteq})
where the operator ${\hat K}\left(z',\sigma'\right)$
has the form 
\begin{equation}\label{Ktot}
{\hat K}\left(z',\sigma'\right)=\frac {1}{8\,\pi^2}\,
\otimes_{k=1}^{N-1}{\hat K}(x_k,\mu_k\,\nu_k)
\otimes{\hat K}^{(1/2)}\left(s,\alpha,\,\beta\right).
\end{equation}
Here, the operator ${\hat K}(x_k,\mu_k,\nu_k)$
has the form of Eq.(\ref{Ksym}) with $\epsilon=1,$ 
and the operator 
${\hat K}^{(1/2)}\left(s,\alpha,\,\beta\right)$
is given by formula~(\ref{Kj}) with $j=1/2$.
Moreover, we have to introduce the marginal distribution 
$w\left({\vec x},\vec\mu,\vec\nu,s,\alpha,\beta,t\right)$ 
describing 
a state of spin-$\frac{1}{2}$ particle which 
depends on the continuous variables
${\vec x}$, discrete spin projection $s$, symplectic reference
frame's labels $\vec\mu$ and $\vec\nu$, 
and Euler angles $\alpha$
and $\beta$.  
Then, for a given Hamiltonian ${\hat H}$ the general 
equation~(\ref{inteq}) takes the form of a Pauli-like equation 
equation
\begin{eqnarray}\label{paulieq}
{\partial_t\,w}({\vec x},{\vec\mu},{\vec\nu},s,\alpha,\beta,t)&=&
\sum_{s'=-1/2}^{1/2}
\int d{\vec X'}d{\vec\mu'}d{\vec\nu'}d\Omega'\,\,
w({\vec x'},{\vec\mu'},{\vec\nu'},s',\alpha',\beta',t)\nonumber\\
&\times&
\Theta({\vec x},{\vec\mu},{\vec\nu},s,\alpha,\beta;
{\vec x'},{\vec\mu'},{\vec\nu'},s',\alpha',\beta')\,,
\end{eqnarray}
where 
\begin{equation}\label{Theta}
\Theta=
\frac {i}{8\,\pi^2}
\langle\langle {\vec x}\,, s\mid
\left[
\otimes_{k=1}^{N-1}{\hat K}(x'_k,\mu'_k\,\nu'_k)
\otimes{\hat K}^{(1/2)}\left(s',\alpha',\,\beta'\right),{\hat H}
\right]
\mid {\vec x}\,, s \rangle\rangle\,.
\end{equation}
The structure of the derived Pauli-like equation for 
probability distributions depends on the
particular tomography schemes we have considered.
Obviously, it would be useful to find the schemes which 
give the simplest form for such
dynamical equation, nevertheless this is a nontrivial 
problem related to the possibility
of finding properly transformed projector (\ref{Pizsig}). 
The latter are
investigated in Ref. \cite{brief}, but for different 
purposes.

\subsection{Limit cases}

We want now to consider two limiting cases of the above 
equation (\ref{paulieq}).

First of all we consider the (one-dimensional) spatial 
case only with free motion
\begin{equation}\label{Hfree}
{\hat H}=\frac{{\hat p}^2}{2}\,.
\end{equation}
The spin part does not contribute since ${\hat H}$ does 
not contain the spin operators,
that is
\begin{eqnarray}\label{deltaspin}
&&\int\frac{d\Omega'}{8\pi^2}\,
w(s',\alpha',\beta',-)
\langle\langle s\mid
{\hat K}^{(j)}(s',\alpha',\beta')
\mid s\rangle\rangle
=\sum_{m_1,m_2=-j}^{j}
D^{(j)}_{s\,m_1}(\alpha,\beta,\gamma)
D^{(j)\,*}_{s\,m_2}(\alpha,\beta,\gamma)\nonumber\\
&&\times\sum_{j_3=0}^{2j}\sum_{m_3=-j_3}^{j_3}\sum_{s'=-j}^{j}
(-)^{m_2-s'}(2j_3+1)^2 W_{s'\,-s'\,0}^{j\,j\,j_3} 
\, W_{m_1\,-m_2\,m_3}^{j\,j\,j_3}
\nonumber\\
&&\times
\int\,\frac{d\Omega'}{8\pi^2}\,w(s',\alpha',\beta',-)
D^{(j_3)}_{0\,m_3}(\alpha',\beta',\gamma')
=w(s,\alpha,\beta,-)\,,
\end{eqnarray}
where $-$ indicates other possible variables.
Then, for what concerns the spatial part, it is 
important to calculate the commutator
between the kernel and the Hamiltonian, given by
\begin{equation}\label{comspace}
\left[e^{i\mu'{\hat q}}e^{i\nu'{\hat p}},{\hat p}^2\right]=
e^{i\mu'{\hat q}}e^{i\nu'{\hat p}}(-2\mu'{\hat p}-\mu'^2)\,.
\end{equation}
Now, one can write
\begin{eqnarray}\label{freeinteq}
{\partial_t \, w}(x,\mu,\nu,t)&=&\frac{i}{4\pi}\int dx' d\mu' d\nu'
w(X',\mu',\nu',t) e^{-iX'+i\mu'\nu'/2}
\nonumber\\
&\times&\int dq
\langle\langle x\mid e^{i\mu'{\hat q}}e^{i\nu'{\hat p}} |q\rangle
\langle q| (-2\mu'{\hat p}-\mu'^2)  \mid x \rangle\rangle
\end{eqnarray}
By using the explicit form for the wave functions 
$\langle q\mid x \rangle\rangle$ (\ref{wf}), 
toghether with the homogeneous property (\ref{hom}),
it is possible to reduce the above equation to a 
very simple form
\begin{equation}\label{freeeq}
{\partial_t\, w}=\mu\,\partial_{\nu}\, w
\end{equation}
which was derived in a different way in Ref. \cite{fou}.

As a second case we study the dynamics of 
spin-$\frac{1}{2}$ degree only.
The Hamiltonian we wish to consider is
\begin{eqnarray}\label{Hspin}
{\hat H}=
\left(\begin{array}{cr}
a&0\\
0&c
\end{array}\right)\,.
\end{eqnarray}
Of course, the spatial degree is not affected, 
so its variables can be disregarded;
this also results from the fact that
\begin{equation}\label{deltaspace}
\frac{1}{2\pi}\int dx' d\mu' d\nu'\,
w(x',\mu',\nu',-)\langle\langle x\mid 
e^{i\mu' {\hat q}} e^{i\nu' {\hat p}} 
\mid x \rangle\rangle
e^{-ix'+i\mu'\nu'/2}=
w(x,\mu,\nu,-)
\end{equation}
In this case the relation between the 
transformed spin state projection 
and the untransformed one is given by 
\begin{equation}\label{sofm}
\mid s\rangle\rangle=
{\hat D}^{(1/2)\,*}_{s,1/2}(\alpha,\beta)|
\mbox{$\frac{1}{2}$}\rangle
+{\hat D}^{(1/2)\,*}_{s,-1/2}(\alpha,\beta)|
\mbox{$-\frac{1}{2}$}\rangle
\,.
\end{equation}
Again, the central task is the calculation of 
the commutator between the kernel 
and the Hamiltonian.
It is easy to see that
\begin{eqnarray}\label{comspin}
&&\langle\langle s\mid
\left[ {\hat K}^{(1/2)}(s',\alpha',\beta'),{\hat H} \right] 
\mid s \rangle\rangle=
(-1)^{-s'}\sum_{j_3=0}^{1}W^{1/2\,1/2\,j_3}_{s'\,-s'\,0}(2j_3+1)^2
\nonumber\\
&&\times
\sum_{m_1\neq m_2,\,-1/2}^{1/2}(-)^{m_2}\sum_{m_3=-j_3}^{j_3}
D^{(j_3)}_{0\,m_3}(\alpha',\beta',\gamma')\,
W^{1/2\,1/2\,j_3}_{m_1\,-m_2\,m_3}
\nonumber\\
&&\times
(-)^{1/2-m_2}(a-c)
D^{(1/2)}_{s\,m_1}(\alpha,\beta,\gamma)
D^{(1/2)\,*}_{s\,m_2}(\alpha,\beta,\gamma)
\,.
\end{eqnarray}
Due to the properties of the Wigner-$3j$ symbols 
we may see that the terms
with $j_3=0,1$, and $m_3=0$ do not give contributions;
moreover, changing the value of $s'$, it changes only 
the sign.
Thus, we will get
\begin{eqnarray}\label{spininteq}
{\partial_t\, w}(\mbox{$\frac{1}{2}$},\alpha,\beta,t)&=&
\int \frac{d\Omega'}{8\pi^2}
\left[ w(\mbox{$\frac{1}{2}$},\alpha',\beta',t)
-w(-\mbox{$\frac{1}{2}$},\alpha',\beta',t)
\right]\nonumber\\
&\times&\frac{3}{2}(a-c)\sin\beta'\sin\beta\sin(\alpha-\alpha')
\end{eqnarray}
and by using the normalization condition it can be rewritten as
\begin{equation}\label{spineq}
{\partial_t\,w}(s,\alpha,\beta,t)=
3(a-c)\sin\beta\int \frac{d\Omega'}{8\pi^2}
\,w(s,\alpha',\beta',t)
\sin\beta'\sin(\alpha-\alpha')
\end{equation}
which is similar to that derived in Ref. \cite{dodman}
(the differencies are due to the degeneracy of the 
spin-$\frac{1}{2}$ systems).
It should be noted in the above equation that the 
argument $s$ is the same
in both sides; this is consistent with the fact 
that ${\hat H}$ in Eq.(\ref{Hspin})
does not mix states with different $s$.
On the other hand it can be easily checked 
that the sum over $s$ at r.h.s. of Eq.(\ref{spineq}) 
causes the integral to become zero;
this is consistent with the fact that at l.h.s. 
we will obtain the time derivative of a
constant. Also, if $a=c$, the r.h.s. of 
Eq.(\ref{spineq}) will be zero since the Hamiltonian 
(\ref{Hspin}) will be proportional to the 
identity and will not produce any evolution.

\section{Examples}

In the previous section, we discussed the probability 
for the joint 
measurement of the spin and spatial variables. 
Therefore, here we would like to 
consider some examples involving both variables.

At first we consider a system with the following hamiltonian
\begin{equation}\label{Htrap}
{\hat H}=\frac{1}{2}\left({\hat p}^2+{\hat q}^2\right)
+\left(|\mbox{$\frac{1}{2}$}\rangle\langle\mbox{$\frac{1}{2}$}|
-|\mbox{$-\frac{1}{2}$}\rangle\langle\mbox{$-\frac{1}{2}$}|\right)
\,.
\end{equation}
It could describe e.g. one vibrational 
degree of a trapped electron plus its spin \cite{bg}.
The measurability of marginals in this system 
is investigated in Ref. \cite{mic}.
Here, as a straigthforward extension of the 
arguments of Sec. VI.1 we obtain
\begin{eqnarray}\label{trapeq}
\partial_t\,w(x,\mu,\nu,s,\alpha,\beta)&=&
\left(\mu\partial_{\nu}-\nu\partial_{\mu}\right)
w(x,\mu,\nu,s,\alpha,\beta)\\
&+&6\sin\beta\int \frac{d\Omega'}{8\pi^2}\,
w(x,\mu,\nu,s,\alpha',\beta',t)
\sin\beta'\sin(\alpha-\alpha')\nonumber
\,.
\end{eqnarray}
Let us now consider an initial entangled state like
\begin{equation}\label{Psiini}
\Psi(0)=\frac{1}{\sqrt{2}}\left(
\mid 0\rangle\otimes\mid-\mbox{$\frac{1}{2}$}\rangle
+\mid 1\rangle\otimes\mid\mbox{$\frac{1}{2}$}\rangle
\right)\,,
\end{equation}
where $\mid n\rangle$ represents the number 
eigenstate of a harmonic oscillator.
At Eq.(\ref{Psiini}) corresponds the following marginal
\begin{equation}\label{wini}
w(x,\mu,\nu,s,\alpha,\beta,t=0)=
\frac{1}{2}\left[
w_{00\downarrow\downarrow}
+w_{11\uparrow\uparrow}
+w_{01\downarrow\uparrow}
+w_{10\uparrow\downarrow}
\right]\,,
\end{equation}
where
\begin{eqnarray}
w_{00\downarrow\downarrow}&=&
\frac{1}{\sqrt{\pi(\mu^2+\nu^2)}}
\exp\left[-\frac{x^2}{\mu^2+\nu^2}\right]
D^{(1/2)\,*}_{s\,-\mbox{$\frac{1}{2}$}}(\alpha,\beta,\gamma)
D^{(1/2)}_{s\,-\mbox{$\frac{1}{2}$}}(\alpha,\beta,\gamma)\,,
\label{wcomp1}\\
w_{11\uparrow\uparrow}&=&
\frac{2x^2}{\sqrt{\pi(\mu^2+\nu^2)^3}}
\exp\left[-\frac{x^2}{\mu^2+\nu^2}\right]
D^{(1/2)\,*}_{s\,\mbox{$\frac{1}{2}$}}(\alpha,\beta,\gamma)
D^{(1/2)}_{s\,\mbox{$\frac{1}{2}$}}(\alpha,\beta,\gamma)\,,
\label{wcomp2}\\
w_{01\downarrow\uparrow}&=&
i\frac{\sqrt{2}x(\nu-i\mu)}{\sqrt{\pi(\mu^2+\nu^2)^3}}
\exp\left[-\frac{x^2}{\mu^2+\nu^2}\right]
D^{(1/2)\,*}_{s\,-\mbox{$\frac{1}{2}$}}(\alpha,\beta,\gamma)
D^{(1/2)}_{s\,\mbox{$\frac{1}{2}$}}(\alpha,\beta,\gamma)\,,
\label{wcomp3}\\
w_{10\uparrow\downarrow}&=&w_{01\downarrow\uparrow}^*
\label{wcomp4}\,.
\end{eqnarray}
Then, the solution of the Pauli equation (\ref{trapeq}) is
\begin{equation}\label{wsol}
w(x,\mu,\nu,s,\alpha,\beta,t)=
\frac{1}{2}\left[
w_{00\downarrow\downarrow}
+w_{11\uparrow\uparrow}
+w_{01\downarrow\uparrow} e^{3it}
+w_{10\uparrow\downarrow} e^{-3it}
\right]\,.
\end{equation}

As a second example, we want to consider the case of 
Landau levels \cite{landau}, i.e.
a charged particle moving in a classical
magnetic field ${\vec B}$ being time-independent and 
axial symmetric. The particle's movement along 
the axis being free, instead  
the Hamiltonian of the transverse motion reads
\begin{equation}\label{H}
{\hat H}=\frac{1}{2}\left[
\left({\hat p}_1-{\hat A}_1\right)^2+
\left({\hat p}_2-{\hat A}_2\right)^2\right],
\qquad {\hat {\vec A}}=\left[{\vec B}\times 
\frac{\hat{\vec r}}{2}\right]\,,
\end{equation}
where ${\hat{\vec r}}=\left({\hat q}_1,\,{\hat q}_2\right)$ 
is the radius-vector of the particle's center,
${\hat p}_1$ and ${\hat p}_2$ are the particle's momentum 
components in the transverse plane.
Having ${\vec B}$ along the third axis and choosing 
$\mid {\vec B}\mid=2$, we get
\begin{equation}\label{Hll}
{\hat H}=\frac{1}{2}\left(
{\hat p}_1^2+{\hat p}_2^2+{\hat q}_1^2+{\hat q}_2^2\right)
+\left({\hat p}_1{\hat q}_2-{\hat p}_2{\hat q}_1\right)
+\left(|\mbox{$\frac{1}{2}$}\rangle\langle\mbox{$\frac{1}{2}$}|
-|\mbox{$-\frac{1}{2}$}\rangle\langle\mbox{$-\frac{1}{2}$}|\right)
\end{equation}

In this case the kernel $\Theta$ of Eq.(\ref{Theta}) is given by
\begin{eqnarray}\label{Thll}
\Theta&=&\frac{i}{4\pi^2}\int\frac{dq_1}{\nu_1}
\int\frac{dq_2}{\nu_2}
e^{i\sum_{l=1}^{2}[(\mu'_l\nu'_l/2-x'_l)
+\mu_l (q_l-\nu'_l)+(x_l \nu'_l+\mu_l {\nu'_l}^2
-\mu_l\nu'_l q_l)/\nu_l]}
\nonumber\\
&\times&\Bigg\{
\sum_{l=1}^{2}(\mu'_l\mu_l q_l/\nu_l-\mu'_l x_l/\nu_l+\nu'_l q_l
-{\mu'_l}^2/2-{\nu'_l}^2/2)
-(x_2-\mu_2q_2)\nu'_1/\nu_2
\nonumber\\
&+&(\mu'_2q_1-\mu'_2\nu'_1)
+(x_1-\mu_1q_1)\nu'_2/\nu_1
-(\mu'_1q_2-\mu'_1\nu'_2)
\Bigg\}\,
\delta_{s,s'}\,\delta(\Omega-\Omega')
\nonumber\\
&+&\frac{6}{8\pi^2}\sin\beta\sin\beta'\sin(\alpha-\alpha')
\delta({\vec x}-{\vec x'})\delta({\vec\mu}-{\vec\mu'})
\delta({\vec\nu}-{\vec\nu'})\,.
\end{eqnarray}

As nontrivial example we also consider here an initial state
which is the entangled superposition
\begin{equation}\label{Psiinill}
\Psi(0)=\frac{1}{\sqrt{2}}
\left[\mid 0\,0\rangle\otimes\mid -\mbox{$\frac{1}{2}$}\rangle
+\mid 1\,0\rangle\otimes\mid \mbox{$\frac{1}{2}$}\rangle
\right]\,.
\end{equation}
It leads to nonfactorisable marginal 
\begin{equation}\label{winill}
w\left({\vec x},{\vec\mu},{\vec\nu},s,\alpha,\beta,t=0\right)
=\frac{1}{2}\left[
w_{0000\downarrow\downarrow}+w_{1010\uparrow\uparrow}
+w_{0010\downarrow\uparrow}+w_{1000\uparrow\downarrow}
\right]\,,
\end{equation}
where
\begin{equation}\label{wcomp}
w_{n_1\,n_2\,n'_1\,n'_2\,,m_1\,m_2}=
w_{n_1\,n_2\,n'_1\,n'_2}\times
D^{(1/2)\,*}_{s\,m_1}(\alpha,\beta,\gamma)
D^{(1/2)}_{s\,m_2}(\alpha,\beta,\gamma)
\,.
\end{equation}
Here, $m_1=-\frac{1}{2}$ ($m_1=\frac{1}{2}$) 
replaces downarrow (uparrow) 
while the spatial part $w_{n_1\,n_2\,n'_1\,n'_2}$ is 
explicitely calculated in the Appendix.

It is now easy to see that the solution of the 
Eq.(\ref{paulieq}) with the
kernel (\ref{Thll}), 
subject to the above initial condition, is
\begin{equation}\label{wsolll}
w\left({\vec x},{\vec\mu},{\vec\nu},s,\alpha,\beta,t\right)
=\frac{1}{2}\left[
w_{0000\downarrow\downarrow}+w_{1010\uparrow\uparrow}
+w_{0010\downarrow\uparrow}e^{3it}
+w_{1000\uparrow\downarrow}e^{-3it}
\right]\,.
\end{equation}

\section{Conclusion}

We conclude that it is possible to obtain an evolution 
equation for
the tomographic probabilities (marginal distributions) of 
an arbitrary tomography
scheme. The main result of our paper is the analog of the 
Pauli equation
for spin-$\frac{1}{2}$ particle. 

The explicit expression for the marginal distribution 
for a trapped particle as well as for
Landau levels has been studied. 
It results that in 
the nonstationary
case they  obey the analog of the Pauli equation. 

The examples
considered demonstrate that the usual problems of conventional
quantum mechanics can be cast into the form in which 
only positive 
probabilities are used to describe quantum states and 
their evolution.
A possible
disadvantage of the approach proposed is a complicated evolution 
equation~(\ref{paulieq}), but,
perhaps, this is the price one ought to pay for the possibility of 
describing quantum objects in terms
of classical probabilities.

Anyway, our argumentations can constitute a step further 
from the Bohr 
position \cite{bohr} about the inapplicability of 
classical modes of
description in the quantum domain.
In fact,
while we belive that quantum mechanics is not classical physics
in disguise, 
we retain (some) classical concepts still applicable against 
counterintuitive notions like
complex statefunctions.

We also belive that the developed classical-like 
formalism could be 
applied to describe quantum mechanical paradoxes,
because usually, if there is a paradox in quantum 
mechanics, there should 
also be a classical one, perhaps worse \cite{mir}.
These aspects will be investigated in a forthcoming 
paper as well as the
extension of the presented approach to the 
relativistic domain \cite{scripta},
in order to find an analog of the Dirac equation.

\section*{Acknowledgments}

S.M. would like to thank the Lebedev Physical 
Institute for the kind hospitality during the first stage 
of this work. O.V.M. thanks the Department of 
Physics and Mathematics
of the University of Camerino for partial support. 
O.V.M. and V.I.M. are grateful to the Russian Foundation 
for Basic Research for the partial support under 
Project No.~99-02-17753.
S.M. and P.T. are grateful to the M.U.R.S.T. for 
the partial support under 
the Project ``Cofinanaziamento".

\section*{Appendix}

The wave function of the particle's coherent state in 
a magnetic field 
${\vec B}$ is~\cite{malkin}
\begin{equation}\label{Psiab}
\Psi_{\alpha,\,\beta}\left(q_1,q_2\right)
=\frac{1}{\sqrt{\pi}}\exp\left\{-
\frac {q_1^2+q_2^2}{2}-\frac {|\alpha|^2}{2}
-\frac{|\beta|^2}{2}-i\alpha\beta+
\beta\left(q_1+iq_2\right)+i\alpha 
\left(q_1-iq_2\right)\right\},
\end{equation}
where $q_1$ and $q_2$ are the particle's 
coordinates and $\alpha$ and $\beta$
complex numbers.

The coherent state~(\ref{Psiab}) is the superposition 
of number states \cite{malkin}
\begin{equation}\label{Psiexp}
\Psi_{\alpha,\,\beta}\left(q_1,q_2\right)=
\exp\left(-\frac{\mid\alpha\mid^2}{2}
-\frac{\mid\beta\mid^2}{2}\right)
\sum_{n=0}^\infty\sum_{n'=0}^\infty\frac {\alpha^n\beta^{n'}
\Psi_{n\,n'}\left(q_1,q_2\right)}{\sqrt{n!n'!}}\,.
\end{equation}

In view of the general relationship between the 
marginal distribution and 
wave function~\cite{men}, we have 
\begin{eqnarray}\label{wwf}
&&w\left(x_1,x_2,\mu_1,\nu_1,\mu_2,\nu_2\right)=
\frac {1}{4\pi^2\mid\nu_1\nu_2\mid}
\nonumber\\
&&\times\left|\int\int\,\exp\left(
\frac{iy_1^2\mu_1}{2\nu_1}-
\frac{iy_1x_1}{\nu_1}
+\,\frac{iy_2^2\mu_2}{2\nu_2}-
\frac{iy_2x_2}{\nu_2}
\right)\Psi_{\alpha\,\beta}
\left(y_1,y_2\right)~dy_1~dy_2\right|^2,
\end{eqnarray}
where parameters $\mu_1,\,\nu_1,\,\mu_2,\,\nu_2$, as usually, 
mark reference
frames; then,  one obtains for the marginal distribution of 
the particle's coherent 
state without spin in a magnetic field the following expression
\begin{eqnarray}\label{wab}
&&w_{\alpha\,\beta}\left(x_1,x_2,\mu_1,\nu_1,\mu_2,\nu_2\right)=
\frac{\exp\left[-|\alpha|^2-|\beta|^2-i\left(\alpha\beta
-\alpha^*\beta^*\right)\right]}{\pi\,\sqrt{
\left(\nu_1^2+\mu_1^2\right)\left(\nu_2^2+\mu_2^2\right)}}\nonumber\\
&&\times\exp\left\{\frac{\left(\nu_1+i\mu_1\right)
\left(i\alpha \nu_1+\beta\nu_1
-ix_1\right)^2+\left(\nu_1-i\mu_1\right)\left(-i\alpha^*\nu_1
\beta^*\nu_1+ix_1\right)^2}{2\nu_1\left(\mu_1^2+\nu_1^2\right)}\right.
\nonumber\\
&&\left.\quad+\,\frac{\left(\nu_2+i\mu_2\right)
\left(\alpha\nu_2+i\beta \nu_2-ix_2\right)^2
+\left(\nu_2-i\mu_2\right)\left(\alpha^*\nu_2-i\beta^*\nu_2+ix_2
\right)^2}{2\nu_2\left(\mu_2^2+\nu_2^2\right)}\right\}.
\end{eqnarray}
Multiplying (\ref{wab}) by $\exp\left(|\alpha|^2+|\beta|^2\right)$
and expanding the expression obtained into the power series, 
we arrive at 
\begin{equation}\label{wabexp}
w_{\alpha\,\beta}\left(x_1,x_2,\mu_1,\nu_1,\mu_2,\nu_2\right)
e^{|\alpha |^2}\,e^{|\beta|^2}=
\sum_{n_1,n_2,n'_1,n'_2=0}^{\infty}
\frac {\alpha^{n_1}(\alpha^*)^{n_2}\beta^{n'_1}(\beta^*)^{n'_2}
w_{n_1\,n_2\,n'_1\,n'_2}}
{\sqrt{n_1!\,n_2!\,n'_1!\,n'_2!}}\,.
\end{equation}
Taking into account the property of the generating 
function for multivariate Hermite 
polynomials~\cite{nova}, namely,
\begin{equation}\label{multiH}
\exp\left\{-\frac {1}{2}{\vec u}M{\vec u}+{\vec u} M{\vec\zeta}\right\}=
\sum_{n_1,n_2,n'_1,n'_2=0}^{\infty}
\frac {\alpha^{n_1}(\alpha^*)^{n_2}\beta^{n'_1}(\beta^*)^{n'_2}}
{n_1!\,n_2!\,n'_1,\,n'_2!}\,
H^{\left\{M\right\}}_{n_1\,n_2\,n'_1,\,n'_2}\left({\vec\zeta}\,\right)\,,
\end{equation}
where the vector ${\vec u}$ has components 
${\vec u}=\left(\alpha,\,\alpha^*,\,\beta,\,\beta^*\right)$,
and comparing (\ref{wabexp}) with (\ref{multiH}), 
we obtain 
\begin{eqnarray}\label{wcompll}
&&w_{n_1\,n_2\,n'_1\,n'_2}\left(x_1,x_2,\mu_1,\nu_1,\mu_2,\nu_2\right)
=\frac {1}{\pi \,\sqrt{
\left(\nu_1^2+\mu_1^2\right)\left(\nu_2^2+\mu_2^2\right)}}\nonumber\\
&&\times\exp\left(-\frac {x_1^2}{\mu_1^2+\nu_1^2}
-\frac {x_2^2}{\mu_2^2+\nu_2^2}\right)
\frac{H^{\left\{M\right\}}_{n_1\,n_2\,n'_1,\,n'_2}
\left({\vec\zeta}\,\right)}
{\sqrt{n_1!\,n_2!\,n'_1!\,n'_2!}}\,,
\end{eqnarray}
where the $4\times 4$ matrix $M$ reads
\begin{equation}\label{M}
M=\left(\begin{array}{cr}
M^{(1)}&M^{(2)}\\
M^{(4)}&M^{(3)}
\end{array}\right)\,.
\end{equation}
The $2\times 2$ matrices $M^{(r)}$ are given by
\begin{eqnarray}
M^{(r)}_{k,l}&=&
\sum_{j=1}^{2}\frac{\nu_j}{\nu_j+i(-)^l\mu_j}
(-)^{j+(r+1)/2}\,\delta_{k+l,{\rm even}}\,,
\quad r=1,3\,,
\nonumber\\[-2mm]
\label{Mr}\\[-2mm]
M^{(r)}_{k,l}&=&
\sum_{j=1}^{2}i\left[
\frac{\nu_j}
{\nu_j+i(-)^l\mu_j}(-)^{l+(j-1)(r/2-1)}+(-)^{l-1}\right]
\delta_{k+l,{\rm even}}\,,
\quad r=2,4\,,
\nonumber
\end{eqnarray}
The argument of the multivariate Hermite polynomials 
${\vec\zeta}=\left(\zeta_1,\,\zeta_1^*,\,\zeta_2,\,\zeta_2^*\right)$
is expressed in terms of the parameters as follows
\begin{eqnarray}
\zeta_1&=&\frac {ix_1}{\sqrt{\mu_1^2+\nu_1^2}}\,
\exp\left(i\tan^{-1}\frac {\mu_2}{\nu_2}\right)-
\frac {x_2}{\sqrt{\mu_2^2+\nu_2^2}}\,
\exp\left(i\tan^{-1}\frac {\mu_1}{\nu_1}\right),
\nonumber\\[-2mm]
\label{zpar}\\[-2mm]
\zeta_2&=&\frac {ix_2}{\sqrt{\mu_1^2+\nu_1^2}}\,
\exp\left(i\tan^{-1}\frac {\mu_1}{\nu_1}\right)-
\frac {x_1}{\sqrt{\mu_2^2+\nu_2^2}}\,
\exp\left(i\tan^{-1}\frac {\mu_2}{\nu_2}\right).
\nonumber
\end{eqnarray}
Taking $n_1=n_2$ and $n'_1=n'_2$ we obtain the marginal distribution 
$w_{n\,n'}(x_1$,$x_2$,$\mu_1$,$\nu_1$,$\mu_2$,$\nu_2)$
for the Landau level states $\mid nn'\rangle$
\begin{eqnarray}\label{wnnp}
&&w_{n\,n'}\left(x_1,x_2,\mu_1,\nu_1,\mu_2,\nu_2\right)\equiv
w_{n\,n\,n'\,n'}\left(x_1,x_2,\mu_1,\nu_1,\mu_2,\nu_2\right)\nonumber\\
&&=\frac {1}{\pi \,\sqrt{
\left(\nu_1^2+\mu_1^2\right)\left(\nu_2^2+\mu_2^2\right)}n!\,n'!}
\,\exp\left(-\frac {x_1^2}{\mu_1^2+\nu_1^2}
-\frac {x_2^2}{\mu_2^2+\nu_2^2}\right)
\,H^{\left\{M\right\}}_{n\,n\,n'\,n'}\left({\vec\zeta}\,\right),
\end{eqnarray}
where $n$ is the main quantum number and $n'-n=l$ 
is the angular momentum
quantum number.

\end{document}